\documentclass[prl,aps,twocolumn,superscriptaddress,preprintnumbers,amsmath,amssymb]{revtex4}
\usepackage[dvips]{graphicx}% Include figure files
\usepackage{dcolumn}% Align table columns on decimal point
\usepackage{bm}% bold math

\begin{document}

\title{Environment-assisted tunneling as an origin of the Dynes density of states}
\author{J. P. Pekola}
\affiliation{Low Temperature Laboratory, Aalto University School of Science and Technology,
P.O.~Box 13500, FI-00076 AALTO, Finland}
\author{V. F. Maisi}
\affiliation{Centre for Metrology and Accreditation (MIKES), P.O. Box 9, 02151 Espoo, Finland}
\author{S. Kafanov}
\affiliation{Low Temperature Laboratory,  Aalto University School of Science and Technology,
P.O.~Box 13500, FI-00076 AALTO, Finland}
\author{N. Chekurov}
\affiliation{Department of Micro and Nanosciences,  Aalto University School of Science and Technology,
P.O.~Box 13500, FI-00076 AALTO, Finland}
%\affiliation{Low Temperature Laboratory, Helsinki University of Technology, P.O.~Box 3500, 02015 TKK, Finland}
\author{A. Kemppinen}
\affiliation{Centre for Metrology and Accreditation (MIKES), P.O. Box 9, 02151 Espoo, Finland}
\author{Yu.~A. Pashkin}
\affiliation{NEC Nano Electronics Research Laboratories and RIKEN Advanced Science Institute, 34 Miyukigaoka, Tsukuba, Ibaraki 305-8501, Japan}
\author{O.-P. Saira}
\affiliation{Low Temperature Laboratory,  Aalto University School of Science and Technology,
P.O.~Box 13500, FI-00076 AALTO, Finland}
\author{M. M\"ott\"onen}
\affiliation{Low Temperature Laboratory,  Aalto University School of Science and Technology,
P.O.~Box 13500, FI-00076 AALTO, Finland}
\affiliation{Department of Applied Physics/COMP,  Aalto University School of Science and Technology,
P.O.~Box 15100, FI-00076 AALTO, Finland}
\author{J. S. Tsai}
\affiliation{NEC Nano Electronics Research Laboratories and RIKEN Advanced Science Institute, 34 Miyukigaoka, Tsukuba, Ibaraki 305-8501, Japan}

\begin{abstract}
We show that the effect of a high-temperature
environment in current transport through a normal
metal--insulator--superconductor tunnel junction can be described by
an effective density of states (DOS) in the superconductor. In the limit of a
resistive low-ohmic environment, this DOS reduces into the
well-known Dynes form. Our theoretical result is supported by experiments in engineered environments.
We apply our findings to improve the performance of a
single-electron turnstile, a potential candidate for a metrological
current source.
\end{abstract}

\maketitle %{\sl Introduction.}
\emph{Introduction}---The density of states (DOS) of the carriers governs the transport rates in a mesoscopic conductor~\cite{blanter2000}, e.g., in a tunnel junction. Understanding the current transport in a junction in detail is of fundamental interest, but it plays a central role also in practical applications,
for instance in the performance of superconducting qubits \cite{qubits}, of electronic coolers and thermometers \cite{giazotto2006}, and of a single-electron turnstile to be discussed in this Letter~\cite{pekola2008}.  When one or both of the contacts of a junction are superconducting, the one-electron rates at small energy bias should vanish at low temperatures because of the gap in the Bardeen-Cooper-Schrieffer (BCS) DOS~\cite{BCS}.  Yet, a small linear in voltage leakage current persists in the experiments~\cite{giazotto2006,oneill2008,rajauria2008b,koppinen2009,jung2009,ralph1995} that can often be attributed to the Dynes DOS, a BCS-like expression with life-time broadening~\cite{dynes1978,dynes1984}.
%Such a leakage is commonly attributed to non-vanishing DOS in the superconductor within the gap \cite{jung2009}, two-electron Andreev current \cite{eiles1993,pothier1994,hergenrother1994,rajauria2008}, non-equilibrium quasiparticles \cite{martinis2009}, or physical imperfections in the junction.
A junction between two leads admits carriers to pass at a rate that depends on the DOS of the conductors, the occupation of the energy levels, and the number of conduction channels in the junction~\cite{ingold1992}. In general, basic one-electron tunneling coexists with many-electron tunneling, for instance co-tunneling in multijunction systems \cite{averin1990}, or Andreev reflection in superconductors \cite{andreev1964,blonder1982}. However, when the junction is made sufficiently opaque, a common situation in practice, only one-electron tunneling governed by the Fermi golden rule should persist. We demonstrate experimentally that the sub-gap current in a high-quality opaque tunnel junction between a normal metal and a superconductor can be ascribed to photon assisted tunneling. We show theoretically that this leads exactly to the Dynes DOS
with an inverse life-time of $e^2k_BT_{\rm env}R/\hbar^2$, where $T_{\rm env}$ and $R$ are the temperature and effective resistance of the environment.%In normal metal leads, the DOS that affects the tunneling rates is essentially constant in the narrow energy window close to the Fermi level, and the current-voltage ($IV$) characteristic is ohmic with no temperature dependence. The picture is different if Coulomb blockade plays a role: the tunneling rates are suppressed in the blockade region exponentially with decreasing temperature of the leads \cite{ingold1992,schoen1998}.

\begin{figure}[ht]\center
\includegraphics[width=0.99\linewidth]{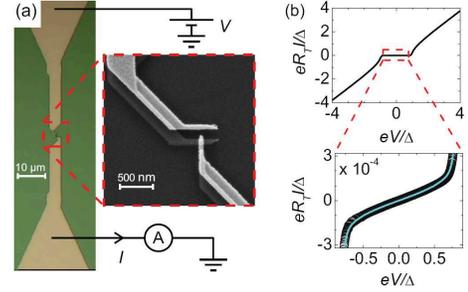}
\caption{\label{fig1} (color online) (a) Geometry of the measured
single NIS junctions made of aluminium (low contrast) as the
superconductor and copper (high contrast) as the normal metal. The
tapered ends lead to $250 \times 250\ \mathrm{\mu m^2}$ pads. (b)
Typical $IV$ characteristics, measured at 50~mK
%base temperature of the cryostat
for a sample with $R_T = 30\ \mathrm{k \Omega}$.
Linear leakage is observed deep in the gap region $|eV| \ll \Delta
\simeq$ 200 $\mu$eV, consistent with the Dynes model using $\gamma =
1.8 \times 10^{-4}$, shown by the cyan line.}
   %The top panel shows a wide bias range with current suppressed within the gap at voltages $|eV| < \Delta \simeq$ 200 $\mu$eV.
\end{figure}

We employ a tunnel junction with a normal
metal--insulator--superconductor (NIS) structure, see
Fig.~\ref{fig1}(a). The essentially constant DOS in the normal metal
renders the NIS junction an ideal probe for the superconductor DOS. Due to the BCS energy gap in an NIS system, the tunneling current is
expected to be exponentially suppressed with decreasing temperature. Yet in
the experiments a small sub-gap current persists as
shown in Fig.~\ref{fig1}(b).
%We note that another possibility to investigate DOS would be a junction between two superconductors; there, however, supercurrent adds to the one-electron current and the interpretation of the results is less obvious than in the case of an NIS junction. %\cite{pekola2004,oneill2008,rajauria2008b,koppinen2009}.
This leakage is typically attributed to Andreev
current~\cite{eiles1993,pothier1994,hergenrother1994,rajauria2008}, smeared DOS of the superconductor~\cite{nahum1993},
non-vanishing DOS in the insulator within the gap~\cite{jung2009},
non-equilibrium quasiparticles~\cite{martinis2009}, or physical
imperfections in the junction. Our junctions, like the one in
Fig.~\ref{fig1}, are made opaque with large normal-state resistance $R_T$
to efficiently suppress the Andreev current. A convenient way to
account for the smearing of the $IV$ characteristics is to use the
so-called Dynes model~\cite{dynes1978,dynes1984} based on an
expression of the BCS DOS with life-time broadening. The Dynes DOS,
normalized by the corresponding normal-state DOS, is given by
\begin{equation}
\label{eq:dynes}
n_S^D(E)=\Big|{\Re \rm e} \big( \frac{E/\Delta+i\gamma}{\sqrt{(E/\Delta+i\gamma)^2-1}} \big) \Big|,
\end{equation}
where $\Delta$ is the BCS energy gap. A non-vanishing $\gamma$
introduces effectively states within the gap region, $|E| < \Delta$,
as opposed to the ideal BCS DOS obtained with $\gamma = 0$ resulting
in vanishing DOS within the gap. This model reproduces the features
observed in our measurements as is shown in Fig.~\ref{fig1}(b). We show that, effectively, the Dynes DOS can be
produced from the ideal BCS DOS by weak dissipative environment at
temperature $T_{\rm env} \gtrsim \Delta / k_B$ promoting photon
assisted tunneling. A similar environment model with comparable
parameter values has also been introduced by other authors to
explain, e.g., observations of excess errors in normal-state
electron pumps~\cite{keller1998,martinis1993} and Andreev reflection
dominated charge transport at low bias voltages in NISIN
structures~\cite{hergenrother1995}.

%Here's some possible citation leftovers for Antti's pool:
%The effect of the electromagnetic environment on NIS-junctions has been considered for the maximum conductivity at the gap edge \cite{anthore2003}. The effect of capacitive shunting has been studied before for Josephson junctions \cite{rimberg1997,steinbach2001}. Other useful ways to engineer the electromagnetic environment include SQUID arrays \cite{watanabe2001} and on-chip resistors \cite{zorin2000, lotkhov2009}. A possible explanation for this photon-assited leakage, dominated by single-electron processes, is $1/f$ noise arising from background charge fluctuations \cite{covington2000,kautz2000}.

\emph{Theoretical results}---For inelastic one-electron tunneling,
the rates in forward ($+$) and backward ($-$) directions through an
NIS junction can be written as
\begin{eqnarray} \label{eq11}
\Gamma_\pm &=&\frac{1}{e^2R_T}\int_{-\infty}^\infty  dE
\int_{-\infty}^\infty  dE'n_S(E')\times \nonumber\\
&&f_N(E\mp eV)[1-f_S(E')]P(E-E'),
\end{eqnarray}
at bias voltage $V$. Here, $P(E)$ refers to the probability density for the
electron to emit energy $E$ to the environment~\cite{ingold1992}. The occupations in the normal and
superconducting leads are given by the Fermi functions
$f_{N/S}(E)=[e^{E/(k_BT_{N/S})}+1]^{-1}$, respectively. In an ideally voltage-biased junction, $P(E) = \delta(E)$.
The current through the junction at low temperature of the leads,
$T_N, T_S \rightarrow 0$, is then
%\begin{eqnarray} \label{eq9}&&
$I^0(V) \equiv e(\Gamma_+^0 - \Gamma_-^0) =\frac{1}{eR_T}\int_{0}^{eV} dE n_S(E)$.
%\end{eqnarray}
Thus we obtain the well-known expression for the conductance of the junction as
\begin{equation} \label{eq10}
{\rm d}I^0/{\rm d}V = R_T^{-1}n_S(eV).
\end{equation}
In view of Eq.~\eqref{eq:dynes}, the non-zero linear conductance at
low bias voltages, typically observed in experiments as shown in
Fig.~\ref{fig1}(b), suggests that the superconductor has
non-vanishing constant density of states within the gap. Within the Dynes model of Eq.~\eqref{eq:dynes}, the normalized DOS at low energies, $|E| \ll \Delta$,
equals $\gamma$,
the ratio of the conductance at zero bias and that at large bias
voltages. Although this approach is correct
mathematically, it is hard to justify the presence of subgap states
physically.

\begin{figure}[t] \center
\includegraphics[width=0.99\linewidth]{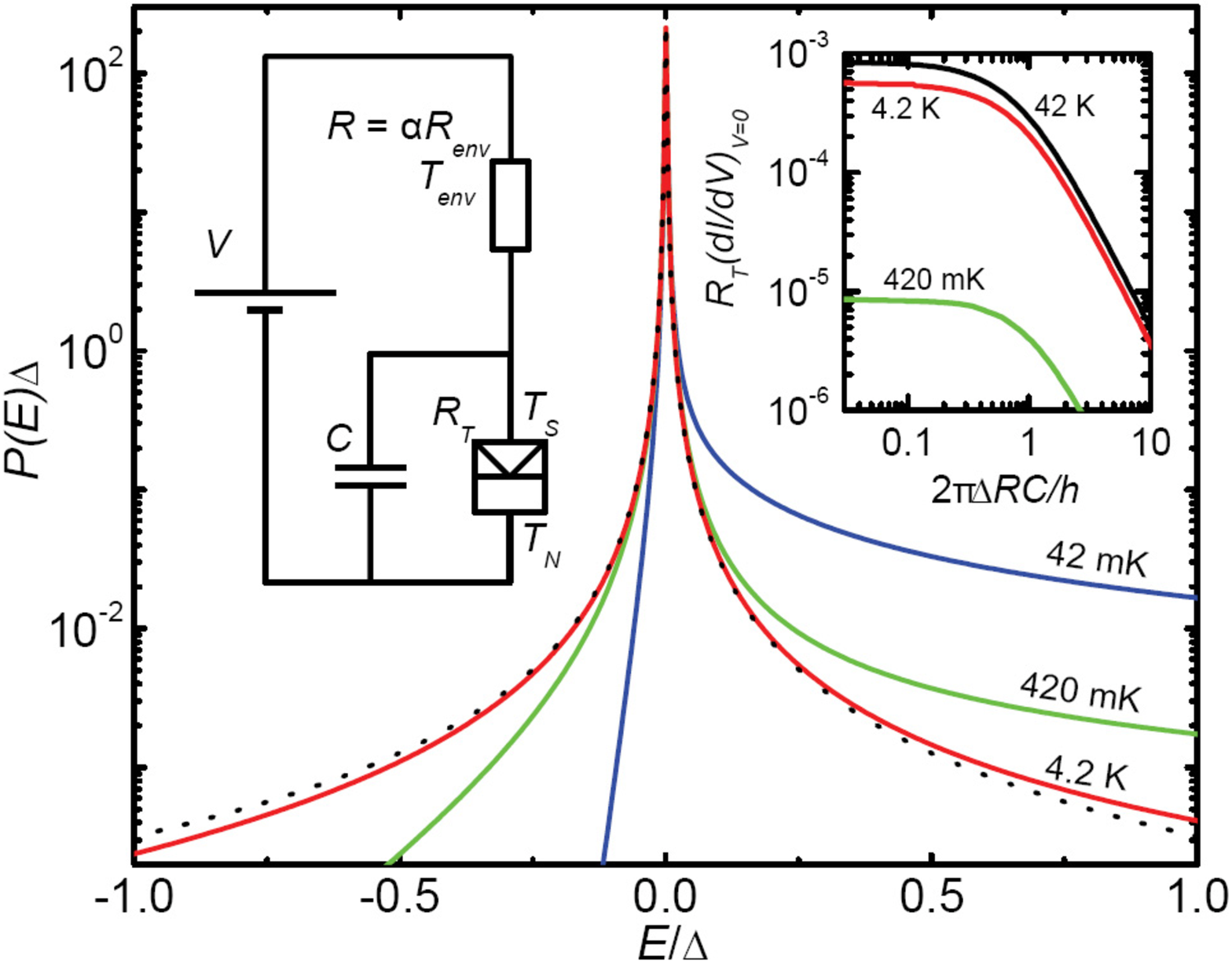}
\caption{\label{fig2} (color online) The probability density $P(E)$
calculated for $\sigma = 10^{-3}= RT_{\rm env}k_B/(R_Q\Delta)$ and
for a few values of the environment temperature, $T_{\rm env}=0.042,
0.42,$ and $4.2$ K. The dotted line is the corresponding Lorentzian
limit of Eq.~\eqref{eq7}. The left inset shows the employed circuit
model of an NIS junction (the rectangular symbol at the bottom
right) in an $RC$ environment. The right inset shows the calculated
zero-bias conductance of the NIS junction as a function of the
capacitance $C$ with the environment corresponding to $\sigma
=10^{-3}$ and $T_{\rm env}=0.42, 4.2$, and $42$~K. We use $\Delta =
200$ $\mu$eV $\simeq k_B\times 2.3$~K for aluminum.}
   %Large scale features of the current $I$ as a function of bias voltage $V$ and capacitance $C$. No dependence on the environment capacitance is visible on this scale. The resistance $R = 2\ \mathrm{\Omega}$ and environment temperature $T_{R} = 4.2\ \mathrm{K}$ are fixed to values consistent with the experiments. {\bf c}, A zoom of {\bf b} which shows the effect of the capacitance. With low $C$, the results match with the analytical ones while with large $C$, leakage is suppressed when ideal voltage bias becomes a better approximation. Solid red and green lines show the best fits to the experimental data of Fig.~\ref{fig3}.
\end{figure}
% Parameters of Fig. 2:
% b: $\gamma_{eff,low C} = 1e-3, T = 4.2 K, 420 mK, 42 mK, \Delta RC/\hbar = 10^{-1}$.
% c: $\gamma_{eff,low C} = 1e-3$.

Here, we base our analysis on the pure BCS DOS ($\gamma=0$) and show
that the Dynes model in Eq.~\eqref{eq:dynes} is consistent with
weakly dissipative environment when Eq.~\eqref{eq11} is used to
obtain the current $I(V)=e(\Gamma_+ - \Gamma_-)$. The effective
resistance value $R = \alpha R_{\rm env}$ of the environment arises
generally from a possibly larger real part of the environment
impedance, $R_{\rm env}$, which is suppressed by a factor of
$\alpha$ due to low-temperature filtering, see Fig.~\ref{fig2}. With this environment,
one obtains the probability density $P(E)$ in the limit of small
$R\ll R_Q\equiv \hbar/e^2$ and for energies $E \ll \hbar(RC)^{-1},
k_BT_{\rm env}$ as a Lorentzian~\cite{online}
\begin{equation} \label{eq7}
P(E)\simeq \frac{1}{\pi \Delta}\frac{\sigma}{\sigma^2+(E/\Delta)^2},
\end{equation}
where $\sigma = Rk_BT_{\rm env}/(R_Q\Delta)$.
%In this approximation, $P(E)$ is symmetric in energy $E$. Yet at high energies the dependence is asymmetric according to the detailed balance, $P(-E) = e^{-E/k_BT_{\rm env}}P(E)$.
As the current of an NIS junction is determined by the values of
$P(E)$ at $|E| \lesssim \Delta$, we can apply Eq.~\eqref{eq7} when
$k_BT_{\rm env} \gtrsim \Delta$, see Fig.~\ref{fig2} for a numerical
demonstration. For a general symmetric $P(E)$ and
$T_N,T_S\rightarrow 0$, one obtains from Eq.~\eqref{eq11} in analogy
with Eq.~\eqref{eq10}:
%\begin{eqnarray} \label{eq9eff}
$I(V) =\frac{1}{eR_T}\int_{0}^{eV} dE n_S^\sigma(E)$,
%\end{eqnarray}
and
\begin{equation} \label{eq10eff}
{\rm d}I/{\rm d}V = R_T^{-1}n_S^\sigma(eV),
\end{equation}
where the effective DOS is given by the convolution
\begin{equation} \label{eq14}
n_S^\sigma(E) \equiv \int_{-\infty}^\infty  dE' n_S(E')P(E-E').
\end{equation}
For the weak resistive environment described by Eq.~\eqref{eq7},
the convolution of a Lorentzian gives
\begin{equation} \label{eq16}
n_{S}^\sigma(E)=\Big | {\Re \rm e}
\big(\frac{E/\Delta+i\sigma}{\sqrt{(E/\Delta+i\sigma)^2-1}}\big)\Big |.
\end{equation}
%$n_S^\sigma$ is thus given by
%\begin{eqnarray} \label{eq16}
%n_{S}^\sigma(E)&=& \frac{1}{\pi}\int_{|\epsilon|\ge 1} d\epsilon
%\frac{ |\epsilon|}{\sqrt{\epsilon
%^2-1}}\frac{\sigma}{\sigma^2+(\epsilon-E/\Delta)^2}\nonumber \\
%&=&\Big |{\Re \rm e}
%\big(\frac{E/\Delta+i\sigma}{\sqrt{(E/\Delta+i\sigma)^2-1}}\big)\Big
%|.
%\end{eqnarray}
This expression is identical to the Dynes DOS in
Eq.~(\ref{eq:dynes}) by setting $\sigma =\gamma$, with the
equivalent inverse lifetime $e^2k_BT_{\rm env}R/\hbar^2$. The
correspondence between the $P(E)$ theory and the Dynes model, our main theoretical
result, is
valid for non-zero lead temperatures as well, as we show in the
supplementary material~\cite{online}. Below, we present numerical and experimental studies
verifying our claim.

\begin{figure}[ht]\center
\includegraphics[width=0.99\linewidth]{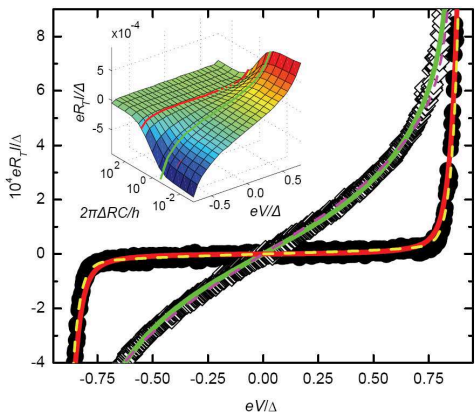}
\caption{\label{fig3} (color online) Measured $IV$ curves of an NIS
junction with $R_T=761$ k$\Omega$ on the ground plane (solid symbols), and of a similar
junction with $R_T=627$ k$\Omega$ without the ground plane (open symbols). Solid lines present the results of the full
$P(E)$ theory for capacitance $C = 10\ \mathrm{pF}$ (red line) and
$C = 0.3\ \mathrm{pF}$ (green line). The resistance and the
temperature of the environment are set to $R = 2\ \mathrm{\Omega}$
and $T_{\rm env} = 4.2\ \mathrm{K}$, respectively, and $\Delta =
200$~$\mu$eV. The dashed lines correspond to the Dynes model with
the parameters yielding the best fit to the data. The normalized
zero bias slope is $5.3\times10^{-4}$ for the green line, and
$2.6\times10^{-5}$ for the red line. The inset shows $IV$ curves
based on the full $P(E)$ calculation as functions of the shunt
capacitance $C$. The red and green lines are reproduced on this
graph from the main figure.}
\end{figure}

%Parameters of Fig. 3:
% a & b:  \Delta_{assumed} = 200 uV = 2.3 K, T_{env} = 4.2 K, \gamma = 9e-4 -> R = 2 \Omega, C_{low} = 0.26 pF (green line), C_{high} = 10.3 pF (red line), \beta_{junction} = 35
% In addition in b: dynes with  $\gamma = 5.3e-4$ (magenta line) and $\gamma = 2.6e-5$ (yellow line).
%Experimental curves: on top of ground plane: $R_T = 761\ k\Omega$, without ground plane: $R_T = 627\ k\Omega$

%{\sl Numerical results.}
\emph{Experiments on single junctions}---The leakage induced by the
electromagnetic environment can be decreased by efficient rf
filtering of the leads and electromagnetic shielding of the sample.
One way to do this without affecting the properties of the junction
itself is to increase the capacitance $C$ across it, see
Fig.~\ref{fig2}. In this way one approaches the case of an ideally
voltage biased junction. In Fig.~\ref{fig2}, we present the
zero-bias conductance of an NIS junction as a function of the
shunting capacitance, $C$, based on the full numerical $P(E)$
calculation. For low $C$, the result using Eq.~\eqref{eq16} is
valid, but for sufficiently high $C$, i.e., for $\Delta RC/\hbar
\gtrsim 1$, the leakage decreases significantly demonstrating that
capacitive shunting is helpful in suppressing the photon assisted tunneling.

%{\sl Experiments.}
To probe the effect of the capacitive shunting in our experiments,
we introduced a ground plane under the junctions.
%We note that this is just one possibility to shunt the junction, however, only on-chip local filters are effective in suppressing the high-frequency noise responsible for the photon assisted tunneling.
The junctions were made on top of an oxidized silicon wafer, where
first a conductive $100\ \mathrm{nm}$ thick $\mathrm{Al}$ layer
working as the ground plane was sputtered. On top of this, a $400\
\mathrm{nm}$ thick insulating high-quality $\mathrm{Al_2O_3}$ film
was formed by atomic layer deposition. The junctions were patterned
by conventional soft-mask electron beam lithography on top of
$\mathrm{Al_2O_3}$. For comparison, junctions were made both with
and without the ground plane.

The experiments reported here were performed in a $^3$He-$^4$He
dilution refrigerator with a base temperature of about 50 mK.  All
the leads were filtered using 1.5~m of Thermocoax cable between the
1~K stage and the sample stage at the base temperature. The $IV$
curves such as the one in Fig.~\ref{fig1}(b) are thermally smeared
at elevated temperatures, but below 200~mK we observe hardly any
temperature dependence. Figure~\ref{fig3} shows the $IV$ curves
measured at the base temperature for one junction on top of a ground
plane and for a similar junction without the ground plane, together
with numerical results from the $P(E)$ theory. The capacitive shunting decreases the zero-bias
conductance significantly. The shunt capacitance values employed in
the $P(E)$ theory, $10\ \mathrm{pF}$ and $0.3\ \mathrm{pF}$,
respectively, match well with the estimates for the experimental
values in each case. The sample without a ground plane with
$C=0.3$~pF is already entering the regime, where the capacitance is
too small to play a role. We used an effective environment
resistance of $R=2$ $\Omega$ at $T_{\rm env}=4.2$ K, close to the values
inferred by Hergenrother et al. \cite{hergenrother1995} in the case
of incomplete shielding. However, the choice of $T_{\rm env}$ is
somewhat arbitrary here: $T_{\rm env}>4.2$ K with correspondingly
lower $R$ would yield a slightly improved fit to the data, but $T_{\rm env}=4.2$ K, the temperature of the outer shield, was chosen as a natural surrounding in the measurement set-up. Our
results with capacitive shunting, on the other hand, correspond to
much improved shielding in the language of Ref.
\cite{hergenrother1995}. Although the experiments of
Ref. \cite{hergenrother1995} are quite different from ours, their
situation resembles ours in the sense that photons with a very high
frequency of $\Delta/h\gtrsim 50$~GHz are responsible for
tunneling.

%We repeated the measurements for several other junctions, with
%similar results. Some of the junctions were fabricated on a
%$\mathrm{SiO_2}$ insulator. For some junctions, we used plasma
%oxidation for the AlO$_x$ tunnel barrier instead of the
%conventional oxidation by admitting a small amount of oxygen to
%the vacuum chamber. This implies that the quality of the tunnel
%barrier and its interfaces can be very good, and that it did not
%affect our results.

% , some of them fabricated on
%a $\mathrm{SiO_2}$ insulator
%, or by plasma oxidation instead of conventional thermal oxidation, %That was not understandable to an outside reader
%with similar results.

\begin{figure}[ht]\center
\includegraphics[width=0.95\linewidth]{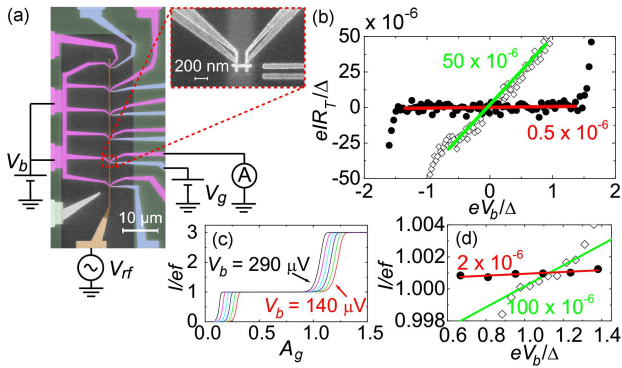}
\caption{\label{fig4} (color online) (a) Scanning electron
micrographs of the SINIS turnstiles. (b) Sub-gap $IV$ curves of
the measured transistors in the gate-open state (charge degeneracy).
The slope of the linear fit corresponds to
the leakage of $0.5\times 10^{-6}$ for the sample with the ground
plane (filled circles) and $50\times 10^{-6}$ for the sample without
the ground plane (open diamonds) in units of the asymptotic
conductance of each SET. (c) Current through the turnstile on the
ground plane as a function of the amplitude of the applied
sinusoidal gate drive at $f=10$~MHz. The gate offset was set to the
charge degeneracy point, and the bias voltage was varied uniformly
between $V_b=140\ldots 290$~$\mu$V. (d) Current at the first
plateau as a function of $V_b$ obtained from data similar
to those in (c) (filled circles) showing leakage of $2\times
10^{-6}$ and for the sample without the ground plane (open diamonds)
with leakage of $100\times 10^{-6}$ and a reduced step width.}
%The inset shows the measured $IV$ curve (grey) when sweeping the gate voltage. The envelopes correspond to the gate open (red) and closed (blue) state, respectively, yielding $R_T=405$~k$\Omega$, $\Delta=210$~$\mu$eV, and charging energy of the SET $E_C\equiv e^2/2C_{\rm tot}=2.5$~K, where $C_{\rm tot}$ is the sum of the junction, gate and stray capacitances.
\end{figure}

\emph{SINIS turnstile}---As a practical application, we discuss the SINIS turnstile
which is a hybrid single-electron transistor (SET) and a strong
candidate for realizing the unit ampere in quantum
metrology~\cite{pekola2008,kemppinen2009,lotkhov2009,kemppinen2009b,maisi2009}.
In the previous experimental
studies~\cite{pekola2008,kemppinen2009,kemppinen2009b}, its accuracy was limited by
the sub-gap leakage. Here we test the influence of the ground plane on the flatness of the current plateaus at multiples of $ef$,
where $f$ is the operating frequency. The ground plane had a 20
$\mu$m wide gap under the SET to reduce the stray capacitance to the
rf gate. The ground plane layer was covered by a 300~nm thick
insulating layer of spin-on glass, on top of which the rf gate and
dc leads were evaporated. Another 300~nm spin-on glass layer was
used to cover the rf gate, and the SET was fabricated on top of this
layer. The device is shown in Fig.~\ref{fig4}(a). This sample
geometry is designed for parallel pumping \cite{maisi2009}, but here
we concentrate on a single device.

Figure~\ref{fig4}(b) shows that in this case, the introduction of
the ground plane reduces the sub-gap leakage by
roughly two orders of magnitude as opposed to a typical turnstile
without the ground plane (the latter data from Ref. \cite{kemppinen2009b}).
%the slope of the linear fit corresponds to an ultralow leakage of
%about $0.5\times 10^{-6}$ in units of the asymptotic conductance of
%the SET. Such a small slope, as compared to the data presented above
%for single junctions, is consistent with the fact that the effective
%resistance and capacitance in the case of an SET are $R/4$ and $4C$,
%respectively, instead of $R$ and $C$ for a single
%junction~\cite{ingold1992}. The corresponding sub-gap $IV$ curve of
%a typical turnstile, but without the ground plane (originally
%reported in \cite{kemppinen2009b}) is shown for reference: the slope
%is about two orders of magnitude higher in this case.
%
%We tested the device performance as a single-electron turnstile.
In the turnstile operation, the current was recorded as a function
of the amplitude of the sinusoidal rf drive, $A_g$, at several bias
voltages. In Fig.~\ref{fig4}(c), we show the quantized current
plateau at $f=10$~MHz, and the averaged current on this plateau is
given in Fig.~\ref{fig4}(d) as a function of the bias voltage. The
differential conductance at the plateau divided by the asymptotic
conductance of the SET is $2\times 10^{-6}$. This result is much improved over those of the earlier
measurements~\cite{pekola2008,kemppinen2009,lotkhov2009}, and that
of the reference sample without the ground plane.
% with about $50$ times higher slope, also shown in Fig.~\ref{fig4}{\bf d}.
%The AC leakage is expected to be lower than at dc because
%the gate is not open most of the time under pumping. In this measurement, however, the situation is opposite. We attribute the difference to local heating due to the rf drive.
%
%The results demonstrate that the ground plane can be used to
%decrease the bias voltage dependence of the SINIS turnstile and
%hence, the accuracy of the device improves. Contrary to the
%resistive environment aiming at the same purpose~\cite{lotkhov2009},
%capacitive shunting does not limit the maximum operation frequency
%fundamentally. However, in these first experiments, we were limited
%to frequencies below the optimum value ($\sim$~100~MHz) set by the
%$RC$ time constant, because of increased on-chip heating which can
%potentially be suppressed by different insulator material and
%geometry of the sample.

%{\sl Conclusions.}
In conclusion, we have shown analytically that the Dynes density of
states can originate from the influence of the electromagnetic
environment of a tunnel junction, and it is not necessarily a
property of the superconductor itself. Our experiments support this
interpretation: we were able to reduce the leakage of an NIS
junction by an order of magnitude by local capacitive filtering. We
stress that capacitive shunting does not necessarily suppress the
subgap leakage of an NIS junction, if the leakage is caused by the
poor quality of the junction, or by true states within the gap due
to, e.g., the inverse proximity effect~\cite{giazotto2006}.
Protecting the junctions against photon assisted tunneling
improves the performance of, e.g., single-electron pumps. Contrary to the
resistive environment aiming at the same purpose~\cite{lotkhov2009},
capacitive shunting does not limit the tunneling rates.

We thank D. Averin, P. Delsing, M. Gustafsson, S. Lotkhov, A.
Manninen, M. Paalanen, and V. Shumeiko for discussions and M.
Meschke, J. Peltonen, and I. Iisakka for technical support. This
work has been supported by Technology Industries of
Finland Centennial Foundation, the Academy of Finland, Emil Aaltonen
Foundation, CREST-JST, MEXT kakenhi "Quantum Cybernetics", and  the European
Community's Seventh Framework Programme under Grant Agreements No.
217257 (EURAMET joint research project
REUNIAM) and No. 218783 (SCOPE).


\begin{thebibliography}{99}
\bibitem{blanter2000} Ya.~M. Blanter and M. B\"uttiker, Phys. Rep. {\bf 336}, 1 (2000).
\bibitem{qubits} {\sl Quantum Computing with Superconducting Qubits}, Quantum Inf. Process. {\bf 8}, pp. 51-281 (2009).
\bibitem{giazotto2006} F. Giazotto {\it et al.}, Rev. Mod. Phys. {\bf 78}, 217 (2006).
%\bibitem{pekola2004} J.~P. Pekola, T.~T. Heikkil\"{a}, A.~M. Savin, J.~T. Flyktman, F. Giazotto, and F.~W.~J. Hekking, Phys. Rev. Lett. {\bf 92}, 056804 (2004).
\bibitem{pekola2008} J.~P. Pekola {\it et al.}, Nature Phys. {\bf 4}, 120 (2008).
\bibitem{BCS} J.\ Bardeen, L.\ N.\ Cooper, and J.\ R.\
Schrieffer, Phys. Rev. {\bf 108}, 1175 (1957).
%\bibitem{mottonen2009} M.\ M\"{o}tt\"{o}nen {\it et al.}, arXiv:0910.0731 (2009).
\bibitem{oneill2008} G. C. O'Neill {\it et al.}, J. Low Temp. Phys. {\bf 151}, 70 (2008).
\bibitem{rajauria2008b} S. Rajauria {\it et al.}, J. Low Temp. Phys. {\bf 153}, 325 (2008).
\bibitem{koppinen2009} P. Koppinen {\it et al.}, J. Low Temp. Phys. {\bf 154}, 179 (2009).
\bibitem{jung2009} H. Jung {\it et al.}, Phys. Rev. B {\bf 80}, 125413 (2009).
\bibitem{ralph1995} D. C. Ralph {\it et al.}, Phys. Rev. Lett. {\bf 74}, 3241 (1995).
\bibitem{dynes1978} R. C. Dynes {\it et al.}, Phys. Rev. Lett. {\bf 41}, 1509 (1978).
\bibitem{dynes1984} R. C. Dynes {\it et al.}, Phys. Rev. Lett. {\bf 53}, 2437 (1984).
\bibitem{ingold1992} G.~L. Ingold and Yu.~V. Nazarov,  in {\em Single charge tunneling}, Vol.~294 of
  {\em NATO ASI Series B}, edited by H. Grabert and M.~H. Devoret (Plenum
  Press, New York, 1992), pp.\ 21--107.
\bibitem{averin1990}  D. V. Averin and Yu. V. Nazarov, Phys. Rev. Lett. {\bf 65}, 2446 (1990).
%\bibitem{averin1992} D.~V. Averin and Yu.~V. Nazarov,  in {\em Single charge tunneling}, Vol.~294 of
 % {\em NATO ASI Series B}, edited by H. Grabert and M.~H. Devoret (Plenum
 % Press, New York, 1992), pp.\ 217--247.
%\bibitem{schoen1998} G. Sch\"{o}n, in {\sl Quantum transport and dissipation}, edited by T. Dittrich, P. H\"{a}nggi, G.-L. Ingold, B. Kramer, G. Sch\"{o}n, and W. Zwerger (Wiley-VCH, Berlin, 1998), Chap. 3.
\bibitem{andreev1964} A.~F. Andreev, Zh. Eksp. Teor. Fiz. {\bf 46}, 1823 (1964);
Sov. Phys. JETP {\bf 19}, 1228 (1964).
\bibitem{blonder1982} G. E. Blonder {\it et al.}, Phys. Rev. B {\bf25}, 4515 (1982).
%\bibitem{hekking1993} F. W. J. Hekking, L. I. Glazman, K. A. Matveev, and R. I. Shekhter, Phys. Rev. Lett. {\bf 70}, 4138 (1993).
%\bibitem{bezuglyi2006} E.~V. Bezuglyi, A.~S. Vasenko, E.~N. Bratus, V.~S. Shumeiko, and G. Wendin, Phys. Rev. B {\bf 73}, 220506 (2006).
%\bibitem{rowell1976} J. M. Rowell and D. C. Tsui, Phys. Rev. B {\bf 14}, 2456 (1976).
\bibitem{eiles1993} T. M. Eiles {\it et al.}, Phys. Rev. Lett. {\bf 70}, 1862 (1993).
\bibitem{pothier1994} H. Pothier {\it et al.}, Phys. Rev. Lett. {\bf 73}, 2488 (1994).
\bibitem{hergenrother1994} J. M. Hergenrother {\it et al.}, Phys. Rev. Lett. {\bf 72}, 1742 (1994).
\bibitem{rajauria2008} S. Rajauria {\it et al.}, Phys. Rev. Lett. {\bf 100}, 207002 (2008).
\bibitem{nahum1993} M. Nahum and J. M. Martinis, Appl. Phys. Lett. {\bf 63}, 3075 (1993).
\bibitem{martinis2009} J.~M. Martinis {\it et al.}, Phys. Rev. Lett. {\bf 103}, 097002 (2009).
%\bibitem{white1986} A. E. White, R. C. Dynes, and J. P. Garno, Phys. Rev. B {\bf 33}, 3549 (1986).
%\bibitem{averin1986} D.~V. Averin and K.~K. Likharev, J. Low Temp. Phys. {\bf 62},  345  (1986).
%\bibitem{likharev1999} K. Likharev, Proceedings of the IEEE {\bf 87},  606  (1999).
%\bibitem{keller1996} M. W. Keller, J. M. Martinis, N. M. Zimmerman, and A. H. Steinbach, Appl. Phys. Lett. {\bf 69}, 1804 (1996).
%\bibitem{nakamura1999} Y. Nakamura, Yu.~A. Pashkin, and J.~S. Tsai, Nature {\bf 398},  786  (1999).
%\bibitem{tinkham} M. Tinkham, {\em Introduction to superconductivity}, 2 ed. (McGraw-Hill, New York, 1996).
\bibitem{keller1998} M. W. Keller {\it et al.}, Phys. Rev. Lett. {\bf 80}, 4530 (1998).
\bibitem{martinis1993} J.~M. Martinis and M. Nahum, Phys. Rev. B {\bf 48},  18316  (1993).
\bibitem{hergenrother1995} J. M. Hergenrother {\it et al.}, Phys. Rev. B {\bf 51}, 9407 (1995).
\bibitem{online} See EPAPS supplementary material for derivation.
\bibitem{kemppinen2009} A.~Kemppinen {\it et al.}, Eur. Phys. J. Spec. Top. {\bf 172}, 311 (2009).
\bibitem{lotkhov2009} S.~V.~Lotkhov {\it et al.}, Appl. Phys. Lett. {\bf 95}, 112507 (2009).
\bibitem{kemppinen2009b} A. Kemppinen {\it et al.}, Appl. Phys. Lett. {\bf 94}, 172108 (2009).
\bibitem{maisi2009} V. F. Maisi {\it et al.}, New J. Phys. {\bf 11}, 113057 (2009).
%Not in use:
%\bibitem{makhlin2001} Y. Makhlin, G. Sch\"{o}n, and A. Shnirman, Rev. Mod. Phys. {\bf 73}, 357 (2001).
%\bibitem{anthore2003} A. Anthore, H. Pothier, and D. Esteve, Phys. Rev. Lett. {\bf 90}, 127001 (2003).
%\bibitem{rimberg1997} A.~J. Rimberg, T.~R. Ho, C. Kurdak, J. Clarke, K.~L. Campman, and A.~C. Gossard, Phys. Rev. Lett. {\bf 78}, 2632 (1997).
%\bibitem{steinbach2001} A. Steinbach, P. Joyez, A. Cottet, D. Esteve, M. H. Devoret, M. E. Huber, and J. M. Martinis, Phys. Rev. Lett. {\bf 87}, 137003 (2001).
%\bibitem{watanabe2001} M. Watanabe and D. B. Haviland, Phys. Rev. Lett. {\bf 86}, 5120 (2001).
%\bibitem{zorin2000} A. B. Zorin, S. V. Lotkhov, H. Zangerle, J. Niemeyer, J. Appl. Phys. {\bf 88}, 2665 (2000).
%\bibitem{covington2000} M. Covington, M. W. Keller, R. L. Kautz, and J. M. Martinis, Phys. Rev. Lett. {\bf 84}, 5192 (2000).
%\bibitem{kautz2000} R. L. Kautz, M. W. Keller, and J. M. Martinis, Phys. Rev. B {\bf 62}, 15888 (2000).

\end{thebibliography}
\end{document}

\title{Environment assisted tunneling as an origin of the Dynes model \linebreak \linebreak On-line material \vspace{3mm}}
%\author{various authors}
\author{J. P. Pekola}
\author{V. F. Maisi}
\author{S. Kafanov}
\author{N. Chekurov}
\author{A. Kemppinen}
\author{Yu.~A. Pashkin}
\author{O.-P. Saira}
\author{M. M\"ott\"onen}
\author{J. S. Tsai}

\date{\today}

%Define labeling for the equations:
\renewcommand{\theequation}{S-\arabic{equation}}
\setcounter{equation}{0}

\maketitle
{\bf TUNNELING RATES IN A RESISTIVE ENVIRONMENT} \vspace*{11pt} \\
Let us consider an NIS tunnel junction with a normal-metal and
a superconducing lead separated by a thin oxide layer, shown in
Fig.~1 of the original article. We assume that the junction is in an
ideal voltage bias $V$ and neglect the charging energy of the single
junction justified by the large capacitance provided by the leads.
As in the fully normal-metal case~\cite{ingold1992}, sequential
tunneling rates $\Gamma_{+/-}$ for single-electron events
along/against the bias are obtained to be
\begin{equation}
\label{eq:tunnelPE} \Gamma_{\pm} =
\frac{1}{e^2R_T}\int^{\infty}_{-\infty} \int^{\infty}_{-\infty} dE
dE'\, n_S(E') f_N(E\mp eV) [1-f_S(E')] P(E-E'),
\end{equation}
where $R_T$ is the asymptotic resistance of the junction, $n_S$ is
the Bardeen-Cooper-Schrieffer (BCS) density of states (DOS) in the
superconductor, and the Fermi functions for the superconductor $f_S$
and the normal metal $f_N$ are governed by the temperatures $T_S$
and $T_N$, respectively. The function $P(E)$ can be interpreted as
the probability density to emit energy $E$ to the environment.
The average electric current through the junction is given by
$I=e(\Gamma_+-\Gamma_-)$. Thus the current--voltage ($IV$) response
of the system, which is typically the only experimental observable
for a single junction, is given by
\begin{equation}
\label{eq:IPE} I(V) = \frac{1}{eR_T}\int^{\infty}_{-\infty}
\int^{\infty}_{-\infty} dE dE'\, n_S(E') [1-f_S(E')][f_N(E -
eV)-f_N(E + eV)] P(E-E').
\end{equation}
We note that in an ideal case with no coupling to an external
electro-magnetic environment in the tunneling process, $P(E)$ is a
delta function and Eqs.~(\ref{eq:tunnelPE}--\ref{eq:IPE}) reduce
into the typical Fermi golden rule results with one-dimensional
integrals. Generally, $P(E)$ can be written as
\begin{equation}
\label{eq:PE} P(E) = \frac{1}{2 \pi \hbar} \int^{\infty}_{-\infty}
\exp \left( J(t) + \frac{i}{\hbar} Et \right) dt,
\end{equation}
and by modeling the environment by an impedance $Z(\omega)$ in a
thermal equilibrium at temperature
$T_\textrm{{\footnotesize{env}}}$, one obtains~\cite{ingold1992}
\begin{equation}
\label{eq:phaPh} J(t) = 2 \int^{\infty}_{0} \frac{d \omega}{\omega}
\frac{\mathrm{Re} [Z(\omega)]}{2\pi R_Q}\left[ \coth \left( \frac{\hbar
\omega}{2k_B T_\textrm{{\footnotesize{env}}}} \right) \left( \cos
\omega t - 1 \right) - i \sin \omega t \right],
\end{equation}
where $R_Q = \hbar/e^2$ is the resistance quantum. For a purely
resistive environment, $\mathrm{Re}[Z(\omega)] = R/[1 + \left(
\omega RC \right)^2 ]$, where $R$ is the resistance of the
environment and $C$ is the total capacitance at the junction
including the junction capacitance and parallel shunt capacitors.
Although this rather simple model has essentially only two free
parameters, $R$ and $T_\textrm{{\footnotesize{env}}}$, for low values of capacitance $C$, it has been
successfully applied to explain several experimental
observations~\cite{Martinis,Tinkham}.

In our case of a resistive environment, the integral in
Eq.~(\ref{eq:phaPh}) can be evaluated to yield
\begin{equation}
\label{eq:phaPhaRC} J(t) = \frac{\rho}{2} \left[ \cot (B) (1-
e^{-\left| \tau \right|} ) - \frac{\left| \tau \right|}{B} -2
\sum_{n = 1}^{\infty} \frac{1-e^{-n \pi \left| \tau \right| / B}} {n
\pi \left[1- \left(n \pi/B \right)^2 \right]} -i\,
\mathrm{sign}\left(\tau \right) (1-e^{- \left| \tau \right|} )
\right],
\end{equation}
where $\rho = R / R_Q$, $\tau = t/(RC)$ and $B =
\hbar/(2k_BT_\textrm{{\footnotesize{env}}}RC)$. For
$|\tau|\gg 1,B$, we can neglect all the terms in
Eq.~(\ref{eq:phaPhaRC}) containing $|\tau|$ in the exponent.
Use of this approximation for evaluating $P(E)$ according to Eq.~(\ref{eq:PE}) is justified if
the energy exchange $E$ with the environment satisfies
$|E|\ll\hbar/(RC),\,k_BT_\textrm{{\footnotesize{env}}}$. %
For low bias voltages in an NIS junction, an electron needs to absorb energy of $\Delta -eV$ in order to tunnel. Therefore, the BCS gap gives the relevant scale of the energy exchange at low bias voltages. At higher voltages, the current is essentially linear in $V$ and insensitive to the functional from of $P(E)$. Hence the conditions above
can be taken to
$\Delta\ll\hbar/(RC),\,k_BT_\textrm{{\footnotesize{env}}}$.
Thus, under these circumstances we
simplify Eq.~(\ref{eq:phaPhaRC}) into
\begin{equation}
\label{eq:phaPhaSmallR} J(t) = -\frac{\rho}{2} \left[ \alpha +
\frac{\left| \tau \right|}{B} + i\, \mathrm{sign}\left(\tau \right)
\right],
\end{equation}
where $\alpha$ is the sum of all terms in Eq.~(\ref{eq:PE})
independent of $\tau$. By evaluating the integral in
Eq.~(\ref{eq:PE}) with $J(t)$ from Eq.~(\ref{eq:phaPhaSmallR}), we
obtain
\begin{equation}
\label{eq:PEsmallRinterm} P(E) = \frac{1}{\pi} e^{-\rho \alpha/2}
\frac{\left( \rho k_B T_\textrm{{\footnotesize{env}}} \right) \cos
\left( \rho/2 \right) + E \sin \left( \rho/2 \right) }{\left( \rho
k_B T_\textrm{{\footnotesize{env}}} \right)^2 + E^2}.
\end{equation}
Here, the normalization condition $\int_{-\infty}^\infty P(E)dE = 1$
yields $e^{-\rho \alpha/2} = 1/\cos \left( \rho/2 \right)$.
Furthermore, the situation $\rho\ll 1$ is met in
experiments~\cite{Martinis,Tinkham} unless the environment in the
immediate vicinity of the junction is deliberately engineered to be
high Ohmic~\cite{Lotkhov}. Together with the assumption $|E|\ll
k_BT_\textrm{{\footnotesize{env}}}$ cast above,
Eq.~(\ref{eq:PEsmallRinterm}) assumes the form
\begin{equation}
\label{eq:PEfinal} P(E) = \frac{1}{\pi}\frac{\left( \rho k_B
T_\textrm{{\footnotesize{env}}} \right)}{\left( \rho k_B
T_\textrm{{\footnotesize{env}}} \right)^2 + E^2}.
\end{equation}
We note that since we neglected the term $E \sin \left( \rho/2
\right)$ in Eq.~(\ref{eq:PEsmallRinterm}), the detailed balance
$P(-E)=e^{-E/(k_BT_\textrm{{\footnotesize{env}}})}P(E)$ is not
perfectly satisfied. Due to our assumption of high temperature
$|E|\ll k_BT_\textrm{{\footnotesize{env}}}$, the error due to the
slight breaking of the detailed balance is, however, small. As
shown below, the fact that the $P(E)$ function is symmetric, as in
Eq.~(\ref{eq:PEfinal}), is crucial in reducing $IV$ characteristics
given by Eq.~(\ref{eq:IPE}) into the typical form with a
one-dimensional integral in the energy domain. In this form, the
effect of the environment is only to effectively modify the BCS DOS.

%Konsistenssi tsekki $\rho k_BT_\textrm{{\footnotesize{env}}}\ll k_BT_\textrm{{\footnotesize{env}}}$ ja $\rho k_BT_\textrm{{\footnotesize{env}}}\ll \hbar/(RC)$ mist? seuraa $\rho\ll 1$ ja $\rho\ll 2B$

\vspace*{11pt}
{\bf EFFECTIVE DENSITY OF STATES}
\vspace*{11pt}

Let us come back to Eq.~(\ref{eq:IPE}) yielding the current--voltage
characteristics of an NIS junction. Since the BCS DOS vanishes inside
the energy gap $|E|<\Delta$, Eq.~(\ref{eq:IPE}) can be expressed in
the form
\begin{equation}
\label{eq:IPEn} I(V) = \frac{1}{eR_T}\int^{\infty}_{-\infty}dE
\int^{\infty}_{0} dE'\, n_S(E')[f_N(E - eV)-f_N(E + eV)] P(E-E'),
\end{equation}
if the temperature of the superconductor satisfies $T_S\ll
\Delta/k_B$. This condition is met in the experiments reported here
since $T_S<100\textrm{ mK}\ll\Delta/k_B\approx 2\textrm{ K}$. We
expand the integral in Eq.~(\ref{eq:IPEn}) as
\begin{eqnarray}
\label{eq:IPEe} I(V) &=& \frac{1}{eR_T}\int^{0}_{-\infty}dE
\int^{\infty}_{0} dE'\, n_S(E')[f_N(E - eV)-f_N(E + eV)] P(E-E') +
\nonumber \\
&& \frac{1}{eR_T}\int^{\infty}_{0}dE \int^{\infty}_{0} dE'\,
n_S(E')[f_N(E - eV)-f_N(E + eV)] P(E-E').
\end{eqnarray}
By negating the integral variables in Eq.~(\ref{eq:IPEe}) and
employing the symmetry of the Fermi function $f(-x)=1-f(x)$, of the
BCS density of states $n_S(-x)=n_S(x)$, and of the $P(E)$ function
$P(-E+E')=P(E-E')$, we obtain
\begin{equation}
\label{eq:IPEeff} I(V) = \frac{1}{eR_T}\int^{\infty}_{0}dE \,
n_\textrm{{\footnotesize{eff}}}(E)[f_N(E - eV)-f_N(E + eV)],
\end{equation}
where we have introduced the effective DOS
\begin{equation}
\label{eq:neff} n_\textrm{{\footnotesize{eff}}}(E) =
\int^{\infty}_{-\infty}dE'\, n_S(E')P(E-E').
\end{equation}
We stress that to obtain Eq.~(\ref{eq:IPEeff}) which corresponds
exactly to current through an ideal NIS junction without coupling to
environment but with a modified DOS, we only assumed that the
superconductor is at low temperature and the environment is at high
temperature. In this case, it is not possible to distinguish only
from the $IV$ characteristics whether the observed DOS arises, e.g.,
from some intrinsic property of the superconductor such as lifetime
broadening, or from the electromagnetic environment described by
$P(E)$. As reported in the original manuscript, we can make this
distinction by engineering the environment thus showing that it gave
the dominating contribution to the effective DOS.

\pagebreak

{\bf EQUIVALENCE OF THE P(E) AND DYNES DENSITY OF STATES}
\vspace*{11pt}

The Dynes DOS~\cite{Dynes1,Dynes2} is given by
\begin{equation}
\label{eq:dynes} n_S^D(E) =
\left|\Re\textrm{e}\left[\frac{E/\Delta+i\gamma}{\sqrt{(E/\Delta+i\gamma)^2-1}}\right]\right|,
\end{equation}
from which the BCS DOS $n_S$ is obtained in the limit $\gamma\to 0$.
Thus $n_S(E \pm i\gamma\Delta)=n_S^D(E)$, where the sign can be
chosen freely. To show that the effective DOS arising from the
environment in Eq.~(\ref{eq:PEfinal}) is equivalent to the Dynes
model with $\gamma=2\pi R k_B
T_\textrm{{\footnotesize{env}}}/(R_K\Delta)$, we write the inverse
Fourier transform of $n_S^D$ as
\begin{equation}
\label{eq:apu} \mathcal{F}^{-1}\left(n_S^D\right) =
\int_{-\infty}^{\infty}n_S[\omega-i\,\textrm{sign}(t)\gamma\Delta]e^{i\omega
t}d\omega/\sqrt{2\pi}=e^{-\gamma\Delta|t|}\mathcal{F}^{-1}\left(n_S\right).
\end{equation}
On the other hand,
\begin{equation}
\label{eq:dynesfinal}
n_S^D=\mathcal{F}\left[\mathcal{F}^{-1}\left(n_S^D\right)\right]
=\mathcal{F}\left[e^{-\gamma\Delta|t|}\mathcal{F}^{-1}\left(n_S\right)\right]=\mathcal{F}\left(e^{-\gamma\Delta|t|}
\right)*n_S=n_\textrm{{\footnotesize{eff}}},
\end{equation}
where $*$ denotes a convolution integral and we have used the
identity
$\mathcal{F}^{-1}(f*g)=\mathcal{F}^{-1}(f)\,\mathcal{F}^{-1}(g)$.
Thus the effective DOS in Eq.~(\ref{eq:neff}) with the environment
given by Eq.~(\ref{eq:PEfinal}) is equivalent to the Dynes model.